\documentclass[preprint,12pt]{elsarticle}

\usepackage[utf8x]{inputenc}
\usepackage{amssymb,amsfonts,amsmath,epsfig}
\usepackage{array}
\usepackage{fullpage}
\usepackage{framed}
\usepackage{subfigure}
\newcommand{\tstep}{\Delta t}

\journal{Journal of Computational Physics}

\begin{document}

\begin{frontmatter}



\title{Convergence of methods for coupling of microscopic and mesoscopic 
reaction-diffusion simulations}


\author[label1]{Mark B. Flegg}
\author[label2]{\hskip 1cm Stefan Hellander}
\author[label1]{\hskip 1cm Radek Erban}

\address[label1]{Mathematical Institute, University of Oxford, 
24-29 St Giles' Oxford OX1 3LB, United Kingdom; \\
e-mails: flegg@maths.ox.ac.uk, erban@maths.ox.ac.uk}
\address[label2]{Department of Information Technology, Uppsala 
Universitet, Box 480, 751 06 Uppsala, Sweden; \\
e-mail: stefan.hellander@it.uu.se}

\begin{abstract}
In this paper, three multiscale methods for coupling of mesoscopic 
(compartment-based) and microscopic (molecular-based) stochastic
reaction-diffusion simulations are investigated. Two of the three 
methods that will be discussed in detail have been previously reported 
in the literature; the two-regime method (TRM) and the compartment-placement 
method (CPM). The third method that is introduced and analysed in this paper 
is the ghost cell method (GCM). Presented is a comparison of sources of 
error. The convergent properties of this error are studied 
as the time step $\Delta t$ (for updating the molecular-based 
part of the model) approaches zero. It is found that the error behaviour
depends on another fundamental computational parameter $h$, the 
compartment size in the mesoscopic part of the model. 
Two important limiting cases, which appear in applications, 
are considered: \par
$\,$(i) $\Delta t \to 0$ and $h$ is fixed; 
\par (ii) $\Delta t \to 0$ and 
$h \rightarrow 0$ such that $\sqrt{\Delta t}/h$ is fixed. \\ 
The error for previously developed approaches (the TRM and CPM) 
converges to zero only in the limiting case (ii), but not in case (i). 
It is shown that the error of the GCM converges in the limiting case (i). 
Thus the GCM is superior to previous coupling 
techniques if the mesoscopic description is much coarser than 
the microscopic part of the model.
\end{abstract}


\end{frontmatter}


\section{Introduction}
\label{introduction}
\noindent
Multiscale stochastic reaction-diffusion methods which use models with 
different levels of detail in different parts of the computational domain 
are applicable to a number of biological systems, including modelling of 
intracellular calcium dynamics \cite{Flegg:2013:DSN}, MAPK pathway 
\cite{Hellander:2012:CMM} and actin dynamics \cite{Erban:2013:MRS}.
In these applications, a detailed modelling approach (which requires 
simulation of trajectories and reactive collisions of individual 
biomolecules) is only needed in a small part of the computational
domain. The main idea of multiscale methods is then simple
to formulate \cite{Flegg:2012:TRM}: we use
a detailed modelling approach in the small subdomain of interest
and a coarser model in the rest of the computational domain.
In this paper, detailed molecular-based (microscopic) models
will be given in terms of Brownian dynamics 
\cite{Andrews:2004:SSC, vanZon:2005:GFR}.
The remainder of the computational domain will be divided into compartments 
and a mesoscopic (compartment-based) model will be used, i.e.
we will simulate the time evolution of the 
numbers of molecules in the corresponding compartments 
\cite{Engblom:2009:SSR, Hattne:2005:SRD}.

There have been a number of approaches developed for coupling different
reaction-diffusion models. They include coupling of mesoscopic 
(compartment-based) models with coarser (mean-field) PDE-based descriptions 
\cite{Flekkoy:2001:CPF,Alexander:2002:ARS,Wagner:2004:HCF,Moro:2004:HMS},
coupling of microcopic (molecular-based) models with mean-field PDEs 
\cite{Geyer:2004:IBD, Gorba:2004:BDS, Franz:2013:MRA}, and
coupling of microscopic and mesoscopic models 
\cite{Flegg:2012:TRM, Flegg:2013:ATM,Hellander:2012:CMM,Klann:2012:HSG}
A successful multiscale algorithm requires an accurate implementation 
of inter-regime transfer of molecules. In this paper, we will study convergence
properties of two algorithms for coupling microscopic and mesoscopic
descriptions which were previously published in the literature:
the two-regime method (TRM) \cite{Flegg:2012:TRM, Flegg:2013:ATM} and
the compartment-placement method (CPM) \cite{Hellander:2012:CMM}.
One of the conclusions of our analysis is that these algorithms do not 
converge in the limit of small time steps and a fixed compartment size. 
Thus, we also propose another approach, the ghost cell method (GCM) 
which is suitable for this parameter regime.

We will consider a reaction-diffusion model in the computational domain 
$\Omega \subset {\mathbb R}^N$ for both $N = 1$ and $N = 3$. We will 
divide $\Omega$ into
two parts, open sets $\Omega_M$ and $\Omega_C$, which satisfy
\begin{equation}
\overline{\Omega_M} \cup \overline{\Omega_C} = \overline{\Omega}
\qquad\quad\mbox{and}\qquad\quad
\Omega_M \cap \Omega_C = \emptyset,
\label{TRMgeometry}
\end{equation}
where an overbar denotes the closure of the corresponding set. 
The microscopic simulation technique is used in $\Omega_M.$ 
Each molecule, $j$, 
in $\Omega_M$ is considered to be a point particle at some location 
in space, $\mathbf{X}_j(t)$, at time $t$, which is updated according 
to discretized Brownian motion, i.e.
\begin{equation}\label{moleculeupdate}
 \mathbf{X}_j(t+\Delta t) = \mathbf{X}_j(t) + 
 \sqrt{2D_j\Delta t} \boldsymbol{\zeta},
\end{equation}
where $D_j$ is the diffusion constant of the $j$-th molecule, 
$\Delta t$ is a small prescribed time step and $\boldsymbol{\zeta}$ 
is a vector containing zero mean, unit variance normally distributed 
random numbers. 

In this paper, we will study the convergence of
multiscale methods in the limit $\Delta t \to 0$. Since the
discretized Brownian motion  (\ref{moleculeupdate}) is only used 
in $\Omega_M$, we have to specify what will be done in the
remainder of the domain, $\Omega_C$, where the mesoscopic model
is used. In this paper, we distinguish the following 
two cases:

\vskip 2mm

{\leftskip 1cm \parindent 1cm
(i) the mesoscopic model is kept fixed in the limit $\Delta t \to 0$;
\par
(ii) the mesoscopic model is refined as $\Delta t$ approaches zero.
\par}

\vskip 2mm

\noindent
The resolution of the mesoscopic model (compartment size) will be denoted by 
$h$. Of particular interest is the error that is caused as a direct result 
of the coupling and thus we will use the parameter $h$ as a measure of 
the compartment size at/on the interface between the two modelling 
subdomains. In the case of regular cubic 
compartments of volume $h^3$, the parameter $h$ is simply the 
length of an edge of each cube. We will
also consider unstructured meshes where the compartment size
$h$ will be suitably generalized. Using $h$, the cases (i)--(ii) can be 
formulated as follows:

\vskip 2mm

{\leftskip 1cm \parindent 1cm
$\,$(i) $\Delta t \to 0$ and $h$ is fixed; \par 
(ii) $\Delta t \to 0$ and 
$h \rightarrow 0$ such that $\sqrt{\Delta t}/h$ is fixed.
\par}

\vskip 2mm

\noindent
Both limits (i) and (ii) are important in applications. We will see 
that the error at the interface $\partial \Omega_M \cap \partial \Omega_C$
of previously developed methods 
\cite{Flegg:2012:TRM, Flegg:2013:ATM,Hellander:2012:CMM}
converges to zero in the limit (ii). This limit requires the
refinement of the mesoscopic model. However, the standard mesoscopic model 
converges in the limit $h \to 0$ only
if the molecules are subject to zero-order or first-order chemical
reactions \cite{Erban:2007:PGS}. It fails to converge when
bimolecular reactions are present \cite{Erban:2009:SMR}. 
This makes the limit (i) attractive in applications.
In Section \ref{GCM}, we introduce the GCM which  
converges in the limit (i).

The paper is organized as follows. In Section \ref{TRM},
we summarize the TRM for coupling of structured mesoscopic meshes  
with microscopic simulations. The methodology for simulation 
of stochastic reaction-diffusion processes on irregular meshes 
and the implementation of the CPM is presented in 
Section \ref{URDME}. The GCM is introduced in Section \ref{GCM}.
Using numerical examples in Section \ref{results}, we compare 
the computational error associated with the TRM with that of the GCM for 
structured meshes and the CPM with the GCM for unstructured meshes. 
We will then discuss the sources of these errors and ways in 
which they may be reduced.  

\section{The two-regime method (TRM)}
\label{TRM}
\noindent
The two-regime method (TRM) \cite{Flegg:2012:TRM, Flegg:2013:ATM} couples 
microscopic and mesoscopic subdomains by careful selection of the jump 
rate over the interface from the mesoscopic compartments and careful 
placement of these molecules into the microscopic domain. To date, the 
TRM has been used with mesoscopic subdomains with regular meshes
\cite{Flegg:2013:DSN,Erban:2013:MRS}.
The advantage of using this technique is that accuracy can be 
gained in `regions of interest' $\Omega_M$ without the need to 
run computationally expensive microscopic simulations over the entire 
domain $\Omega$. In this section we will briefly cover the two different 
simulation paradigms and then discuss how these paradigms are 
combined using the TRM.

\subsection{Microscopic simulation}

\noindent
The defining characteristic of `microscopic' simulation techniques for 
diffusion is that each molecule in the system is simulated individually 
on a continuous domain. In particular, these techniques follow the 
trajectory of each Brownian molecule to a resolution dependent 
on the time steps that are used. For illustrative purposes we 
consider here a time-driven microscopic algorithm. That is, 
an algorithm with a defined constant time step. Furthermore, 
we will not be considering volume exclusion effects in this manuscript. 
Each molecule, $j$, is therefore considered to be a point particle at 
some location in space, $\mathbf{X}_j(t)$, at time $t$. The Brownian 
diffusion of these molecules is modelled by $(\ref{moleculeupdate})$.
Reactions may take place between these diffusing molecules 
at a particular time step if the reactants are within a given 
reaction radius of each other \cite{Rice:1985:DLR,Lipkova:2011:ABD}. 

Molecule interactions with boundaries depend on the type of 
boundary: boundaries can be reflective, adsorbing or
reactive (partially adsorbing) \cite{Erban:2007:RBC}. 
Considering that $\sqrt{D\Delta t}$ is much smaller than 
the local radius of curvature of the boundary, then the
boundary is locally flat on the 
scale of relative motion of the molecules in one time step.
In the case of absorbing boundaries, molecules are removed 
from the system when they are updated to a position outside 
of the boundary. Since we simulate Brownian motion using
a finite time step, we have to take into account
that a molecule can interact with the boundary
during the time step $[t,t+\Delta t]$ even if its
computed position at time $t+\Delta t$ is inside the simulation
domain. The probability, $P_m$ that this molecule-boundary 
interaction occured within the time interval $(t,t+\Delta t]$ 
is dependent on the diffusion constant and the initial and 
final normal distances from the boundary the molecule is found 
($\Delta x_i$ and $\Delta x_f$ respectively)   
\begin{equation}\label{PM}
 P_m(\Delta x_i,\Delta x_f,D,\Delta t) 
 = \exp\left( \frac{-\Delta x_i\Delta x_f}{D\Delta t}\right).
\end{equation}
This probability will also be important when it comes to coupling 
of microscopic simulations with mesoscopic simulations via 
an interface in the two-regime method
\cite{Flegg:2012:TRM,Flegg:2013:ATM}.

\subsection{Mesoscopic simulation}

\noindent
Mesoscopic approaches to reaction-diffusion processes are simulated 
on a lattice. For the purposes of the TRM we will describe how this 
is done for a regular cubic lattice. The distance between each node 
is $h$. In a mesoscopic model, molecules can be thought to exist 
only at lattice nodes rather than existing in continuous space.
The state of the simulation at any moment of time is defined by 
a set of numbers describing the copy numbers $\mathcal{N}_{i,j}$ 
of molecules of the $i$-th type at the $j$-th lattice point. 
Considering the diffusion of (non-reacting) molecules, the expected 
state of the system $\mathrm{E}(\mathcal{N}_{i,j})$ 
is described by the equation:
\begin{equation}
\label{equationformean}
\frac{d \mathrm{E}(\mathcal{N}_{i,j})}{d t} = 
D_i\sum_k 
\left( q_{k,j}\mathrm{E}(\mathcal{N}_{i,k}) - q_{j,k}\mathrm{E}(\mathcal{N}_{i,j})\right),
\end{equation} 
where $q_{k,j}$ is the propensity per molecule to go from the $k$-th compartment 
to the $j-$th compartment. It is possible to show that for a regular lattice 
with spacing $h$,
\begin{equation}\label{equationforq}
 q_{k,j} = 
\begin{cases}
D_i/h^2, \quad &\text{if }k \text{ and } j\text{ are adjacent lattice points,} \\
0 , \quad &\text{if }k \text{ and } j\text{ are not adjacent lattice points,}
\end{cases}
\end{equation}
results in the recovery of the discretized form of the diffusion partial differential 
equation and can therefore be used to approximate a diffusion process on the lattice correct 
to order $h^2$. The simulation of a mesoscopic reaction-diffusion process usually makes 
use of event-driven algorithms, such as the Gillespie algorithm \cite{Gillespie:1977:ESS}
or the Gibson-Bruck algorithm \cite{Gibson:2000:EES}.
We shall conceptualize the mesoscopic simulation by considering 
that when a molecule is at a particular lattice point, rather than existing at the node, 
it is somewhere at random inside the compartment belonging to the node defined by the 
lattice dual mesh \cite{Engblom:2009:SSR}. That is, for a regular cubic lattice with 
node spacing $h$, each molecule which is at a particular lattice point is thought to 
exist inside the cubic compartment of side length $h$ for which the lattice point is 
at the center. It is important to note that the state of the molecule has no specific 
location but rather is thought to exist in a probabilistic sense uniformly over its 
compartment. 

\subsection{Interfacing microscopic and mesoscopic simulations}

\begin{figure}
\begin{center}
 \includegraphics[scale=0.65]{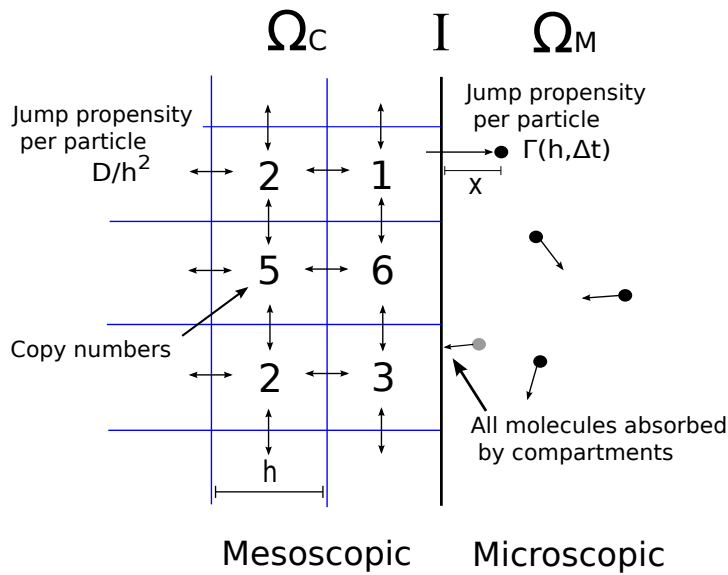}
\caption{{\it Graphical representation of the TRM on a regular square lattice.}}
\label{TRMdiagram}
\end{center}
\end{figure}

\noindent
Interfacing microscopic and mesoscopic simulations of reaction-diffusion processes 
using the TRM has previously been derived for mesoscopic regimes that use regular 
cubic lattices \cite{Flegg:2012:TRM, Flegg:2013:ATM}. The TRM is proposed by partitioning 
the domain $\Omega$ into subdomains (\ref{TRMgeometry}) separted by the interface
$I = \partial \Omega_M \cap \partial \Omega_C$. In both subdomains, molecules behave 
as they would normally according to the rules of that particular regime. We describe 
the TRM with an event-driven mesoscopic simulation in $\Omega_C$ and a time-driven 
microscopic simulation with constant time step $\Delta t$ in
$\Omega_M$. Reactions do not cause any issue within the domain because they occur locally. 
We focus, therefore, on the correct manner in which molecules may migrate over 
the interface $I$. It is assumed that the TRM is simulated such that $\sqrt{D \Delta t}\sim h\ll 1$. 
A diagram of the numerical TRM scheme using a regular cubic lattice can be seen in two 
dimensions in Figure \ref{TRMdiagram}. A detailed TRM algorithm may be found in the 
reference \cite{Flegg:2012:TRM}. In order that a molecular migration over the interface 
is smooth with optimally small error, the propensity $\Gamma$ per molecule to cross the 
interface $I$ from each adjacent compartment is dependent on the parameters $h$ and $\Delta t$. 
For a regular cubic mesoscopic lattice,
\begin{equation}\label{Gamma}
\Gamma(h,\Delta t) = 2\sqrt{\frac{D}{\pi \Delta t h^2}},
\end{equation}
where $D$ is the diffusion constant of the migrating molecule. The TRM considers 
that microscopic molecules in $\Omega_M$ cease to be microscopic molecules, in principle, 
when they migrate over the interface. Molecules are therefore absorbed 
by the interface $I$ from $\Omega_M$ and placed in the closest compartment in $\Omega_C$. 
Equation (\ref{PM}) is used to absorb all molecules which interacted with the
interface. If this is not used then molecules effectively migrate into $\Omega_C$ and back 
out again without changing from a microscopic molecule to a mesoscopic one. 
This is crucial for coupling of the two regimes as outlined in the derivation in 
the reference \cite{Flegg:2012:TRM}. Furthermore, molecules must be precisely placed 
in $\Omega_M$ when migrating from $\Omega_C$. In particular, the perpendicular distance 
$x$ the molecule is placed from the interface into $\Omega_M$ is found by sampling 
from the distribution $f(x)$
\begin{equation}\label{distribution}
 f(x) = \sqrt{\frac{\pi}{4D\Delta t}}\mathrm{erfc}\left( \frac{x}{\sqrt{4D \Delta t}}\right),
\end{equation}
where $\mathrm{erfc}\left( x \right) = 2/\pi \int_x^\infty \exp(-t^2) dt$ is the complementary 
error function. In higher dimensions, the initial position of molecules migrating into 
$\Omega_M$ can be chosen to be uniformly distributed tangentially to the interface 
in the region of the originating compartment \cite{Flegg:2013:ATM}. Then
the error associated with the TRM is $O(h)$. We shall 
investigate the error associated with the TRM in 1D in a later section of this manuscript 
and compare it with the GCM method introduced in Section \ref{GCM}.

\section{Compartment-placement method (CPM)}\label{URDME}

\noindent
In this section, we will discuss how mesoscopic simulation is implemented on an irregular 
lattice \cite{Engblom:2009:SSR}. We will then present a brief description of 
the CPM \cite{Hellander:2012:CMM}. 

\subsection{Mesoscopic simulation on unstructured meshes}

\noindent
Mesoscopic simulations on Cartesian meshes are convenient in the sense 
that they are memory lenient. However, complex geometries and surfaces 
with high curvature, are easier to resolve accurately with an unstructured mesh. 
Living cells can have different shapes and eukaryotes have a complicated internal 
structure with two-dimensional membranes and a one-dimensional 
cytoskeleton \cite{Alberts:2007:MBC}. The geometrical flexibility of 
unstructured meshes is therefore an advantage when considering simulations 
of realistic biological problems.

\begin{figure}[t]
\begin{center}
 \includegraphics[scale=0.65]{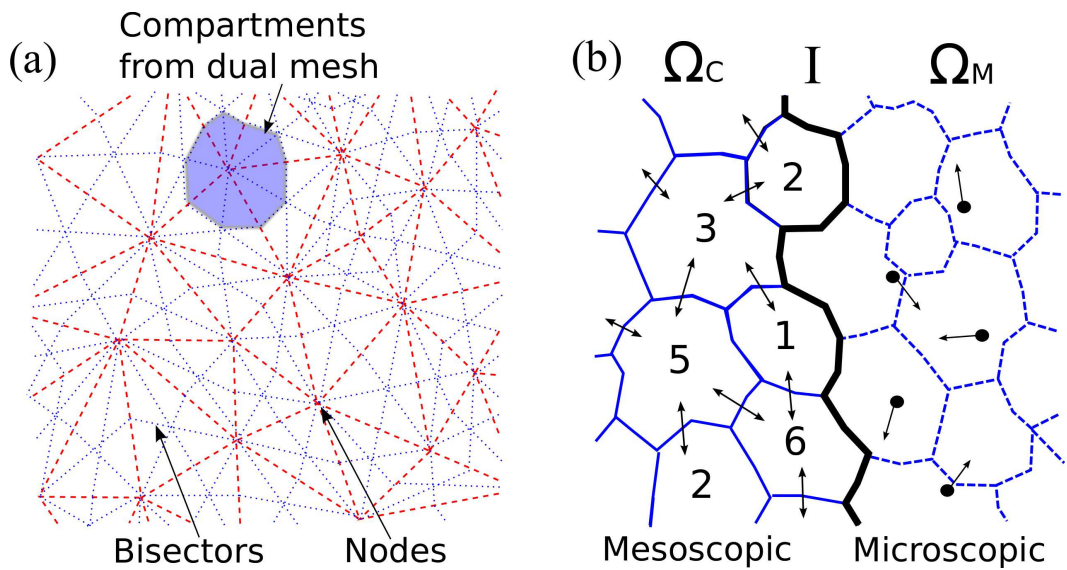}
 \caption{{\it Schematic of CPM computational domain.} 
(a) {\it The primal mesh is indicated with red dashed lines. 
The nodes are connected to form triangles. The bisectors are then drawn 
in to create the dual mesh (blue dotted lines). Compartments are drawn 
from the dual mesh with one node at the center of each compartment. 
One example compartment is shown in blue.} \hfill\break
(b) {\it The domain is split into mesoscopic $\Omega_C$ and microscopic 
$\Omega_M$ domains. Jumps between compartments and from the compartments into 
$\Omega_M$ are calculated using a finite element discretization of the Laplacian. 
The copy numbers of molecules in each compartment in $\Omega_C$ are stored whilst 
in $\Omega_M$ each molecule has its own position in continuous space.}}
\label{compartments}
\end{center}
\end{figure}

Consider a domain $\Omega$. The domain is covered by a primal mesh, such 
that the boundary $\partial\Omega$ is covered with non-overlapping triangles 
and the domain $\Omega$ is covered with non-overlapping tetrahedra 
(resp. triangles in 2D). A dual mesh is constructed from the primal mesh, 
see Figure \ref{compartments}, from the bisectors of the tetrahedra 
(resp. triangles) that use the nodes as vertices. The diffusion of molecules 
is now modelled as discrete jumps between the nodes of the dual mesh. 
The rate $q_{i,j}$ at which a molecules jump from voxel $V_i$ to $V_j$ 
is given by the diffusion constant of the molecule and the finite element 
discretization of the Laplacian on the primal mesh. For details on how the 
unstructured meshes and the diffusion matrix are generated the reader is 
referred to \cite{Engblom:2009:SSR}.

\subsection{Interfacing microscopic and mesoscopic simulations}

\noindent
The algorithm for the CPM is presented in a similar way to the TRM. 
The algorithm progresses asynchronously by updates in the mesoscopic 
simulation and microscopic simulation separately \cite{Hellander:2012:CMM}. 
The jump rates from compartments that are on the interface $I$ between 
regimes are calculated from the underlying mesh over the entire domain. 
That is, the jump rates are calculated by computing the mesoscopic jump 
rates between interfacial compartments and ``compartments'' that are adjacent 
to the interface in the microscopic domain $\Omega_M$ 
(see Figure \ref{compartments}).

Molecules that start in a compartment in $\Omega_C$ and, at the end of 
the time step, have ended up in $\Omega_M$ are initialized uniformly 
inside the ``compartment'' which they jump into, and is the process 
from which the CPM has been named. Molecules in $\Omega_M$ migrate back 
to $\Omega_C$ via microscopic domain diffusion (\ref{moleculeupdate}). 
When a molecule appears inside one of the mesoscopic compartments from 
Brownian motion, it is encorporated into that compartment by increasing 
the copy number inside this compartment.

The CPM has been determined using heuristics. Molecules that are in 
compartments obey mesoscopic rules for diffusive migration. This includes 
molecules that are on interfacial compartments. They jump to compartments 
in $\Omega_M$ as though they were still in $\Omega_C$. When this occurs, 
initialization of the molecules must take place. The molecules are initiated 
uniformly over the compartment in which they are placed. Molecules are not 
placed at the node at the center of this compartment because this would 
unphysically concentrate molecules at this point and reactions would 
occur between possible reactants apon migration over the interface. 
Conversely, molecules that are in $\Omega_M$ obey microscopic rules 
for diffusive migration (Brownian trajectory). When this Brownian trajectory 
leads to a compartment, it can no longer be described using the microscopic 
description and is added to the compartment in which it lands. As we 
shall see, this heuristic approach can lead to inaccuracies.
The inaccuracies can be minimized if $h^2\sim D\Delta t$ 
(that is, if the size of the compartment is approximately the size of 
a microscopic molecular jump).

\section{The ghost cell method (GCM)}
\label{GCM}

\noindent
Here we will consider a new method for interfacing mesoscopic and 
microscopic simulations. This method uses different assumptions to 
the TRM and CPM and is therefore implemented differently. We call 
this method the ghost cell method (GCM) since microscopic molecules 
in $\Omega_M$ feel the presence of a pseudo-compartment allowing for 
instantaneous jumping from $\Omega_M$ to $\Omega_C$ in the same way 
that molecules within compartments jump instantaneously. The steps
of the GCM are given in Table \ref{alg:complete}.

\begin{table}[t]
    \begin{framed}
    \begin{enumerate}
    \setcounter{enumi}{0}
       \item[\mbox{[G.1]}] Initialize lattice over whole domain $\Omega$ and construct dual 
       mesh (compartments). Generate interface $I$ on the edges of compartments 
       to separate $\Omega_M$ from $\Omega_C$. Choose $\Delta t$ and set time $t=0$. 
       Determine $q_{k,j}$ using finite element method between all compartments \cite{Engblom:2009:SSR}.
       \item[\mbox{[G.2]}] Initialize the initial state of the system by placing molecules 
       in compartments in $\Omega_C$ and placing molecules in continuous space in $\Omega_M$. 
       Count and store numbers of molecules in ghost cells, those compartments in $\Omega_M$ 
       which are adjacent to the interface $I$.
       \item[\mbox{[G.3]}] Determine the time $\tau$ for the next event (reaction or diffusive) in 
       $\Omega_C$ or diffusive jumps to and from ghost cells and $\Omega_C$.
       \item[\mbox{[G.4]}] If $t+\tau < \Delta t + \Delta t\lfloor t/\Delta t\rfloor$ then change 
       the state of the system to reflect the next event corresponding to $\tau$ and 
       update time $t:=t+\tau$. If this event is a diffusive jump from ghost cell to $\Omega_C$ 
       choose a molecule at random within the relavant ghost cell to migrate. If this event 
       is a diffusive jump from $\Omega_C$ to a ghost cell then initialize this molecule 
       with uniform probability over the ghost cell.
       \item[\mbox{[G.5]}] If $t+\tau \geq \Delta t + \Delta t\lfloor t/\Delta t\rfloor$ then update 
       the positions of all molecules in $\Omega_M$ using (\ref{moleculeupdate}). Check 
       for reactions in $\Omega_M$ \cite{Flegg:2013:ATM}. All molecules incident on 
       the interface $I$ are reflected. Update time $t:=\Delta t + \Delta t\lfloor t/\Delta t\rfloor$
       \item[\mbox{[G.6]}] Repeat steps [G.3]--[G.5] until the desired end of the simulation.
    \end{enumerate}
    \end{framed}
    \caption{The ghost cell method algorithm.}
    \label{alg:complete}
    \end{table}

The key assumption that is used in the TRM and CPM is that molecules in $\Omega_M$
migrate via diffusion (\ref{moleculeupdate}) over the interface $I$ whereby they
become parts of the corresponding compartment. 
In the GCM, this assumption is relaxed. Instead, molecules migrate over the interface 
using the compartment-based approach in both directions. Microscopic molecules in $\Omega_M$ 
near the interface feel the presence of a layer of ``ghost'' cells (compartments).
In the step [G.2] in Table \ref{alg:complete}, we calculate the numbers of molecules
in these ``ghost'' cells. They are used in the step [G.4] to create a fully compartment-based
simulation of transition across the interface $I$.

\begin{figure}
\begin{center}
 \includegraphics[scale=0.65]{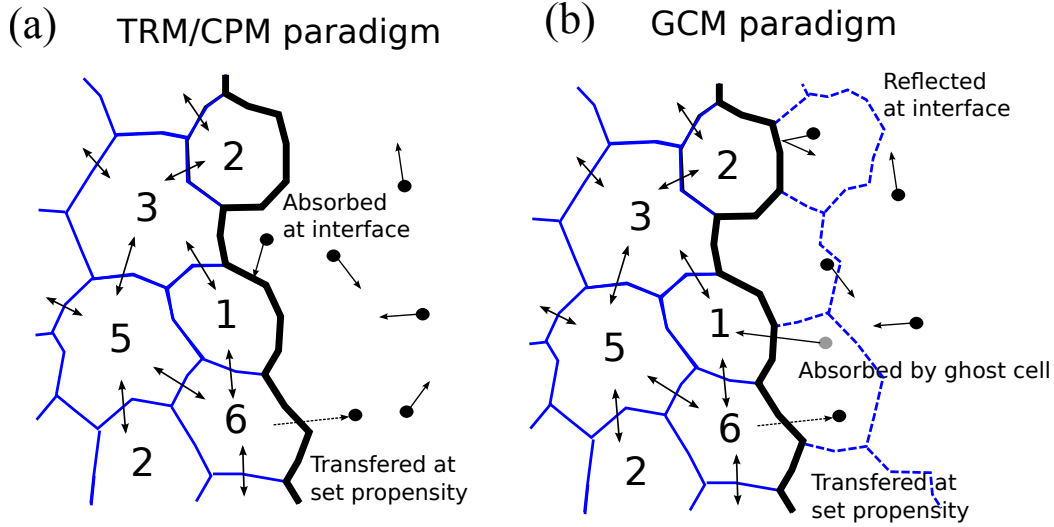}
\caption{{\it A diagram illustrating the fundamental differences between}
(a) {\it the TRM/CPM paradigm and} \hfill\break (b) {\it the GCM paradigm.}}
\label{RegimeDiagram}
\end{center}
\end{figure}

To justify the GCM, let us consider a simulation of diffusion in a domain $\Omega$ for which 
a mesoscopic method was implemented. Then consider the same domain where a microscopic 
simulation is implemented. 
Let the molecules of the microscopic simulation be ``binned'' according to compartments 
of the mesoscopic simulation. The expected number of molecules binned into each 
compartment should match that of the mesoscopic simulation to within the precision 
of the mesoscopic method. This is because both simulations are accurate representations 
of the same phenomena, diffusion. This is the philosophy behind the GCM. Molecules which 
are binned into ghost compartments near the interface may jump into compartments in 
$\Omega_C$ via the rates prescribed by the mesoscopic algorithm. If both regimes are 
correct individually then the flux over the interface $I$ is the same as though 
a mesoscopic algorithm was used over the whole domain. To ensure that microscopic 
molecules do not migrate to $\Omega_C$ via diffusion (\ref{moleculeupdate}), they 
are reflected at the interface $I$ in the step [G.5]. Figure \ref{RegimeDiagram} 
demonstrates the principle 
differences between a TRM/CPM and a GCM description of the interface. In \ref{GCMproof} we 
provide a mathematical analysis of the GCM in one dimension to demonstrate that the expected 
concentration and flux of molecules over the interface are matched. The theoretical error 
associated with the GCM scales as $\sqrt{\Delta t}$ which is on the same order as that of 
the TRM. Unlike the TRM, this error, as we will see in the later part of this manuscript, 
is reduced to zero by reducing $\sqrt{\Delta t}/h$.

The ghost cell method is implemented using the algorithm in Table \ref{alg:complete}. 
This algorithm is given for an event-driven mesoscopic simulation and a time-driven 
microscopic simulation, however it can also be extended to event-driven microscopic 
simulations.

\section{Numerical results and discussion}
\label{results}

\noindent
In this section, we present numerical examples comparing the TRM, CPM and GCM. First,
we demonstrate how the error associated with the interface $I$ is dependent on choices 
of mesh spacing $h$ in the mesoscopic subdomain at the interface and the time step 
chosen for the microscopic subdomain $\Delta t$ for both the TRM and the GCM 
using one dimensional simulations.

\subsection{One dimensional simulations: TRM versus GCM}

\noindent 
We use a simple diffusion test problem to compare the diffusive flow over the 
interface with an exact solution which can be analytically obtained. We use 
the domain $\Omega = (0,1)$ and subdomains $\Omega_C = (0,0.5),$ 
$\Omega_M = (0.5,1)$, which are separated by the interface $I = \{0.5\}$. 
We initially position $N_0 = 5\times 10^5$ molecules according to the distribution 
$g(x) = 2x$, $x\in\Omega$. We construct regular spaced compartments 
of width $h_0 = 0.1$ within $\Omega_C$ and ``bin'' the molecules generated in 
$\Omega_C$ into these compartments. We allow these molecules to diffuse 
throughout the domain $\Omega$ with a diffusion constant $D=1$ using 
the TRM or GCM until time $t=1$. At the boundary $x=0$ molecules 
are absorbed and placed at $x=1$. At the boundary $x=1$ molecules are 
reflected. In this way, $N_0 \, g(x)$ is the steady state distribution
of this system and $0.25 N_0$ is the steady state number of molecules
in $\Omega_C.$ We define a measure of the error $E$ to this test 
problem for each simulation scheme 
\begin{equation}\label{error}
 E = \frac{\sum_j\mathcal{N}_j(1) - 0.25N_0}{N_0},
\end{equation}
where $\mathcal{N}_j$ is the copy number of molecules in the $j$-th compartment 
evaluated at $t=1$ and the sum is taken over all compartments in $\Omega_C$. 

\begin{figure}
\begin{center}
\includegraphics[scale=0.8]{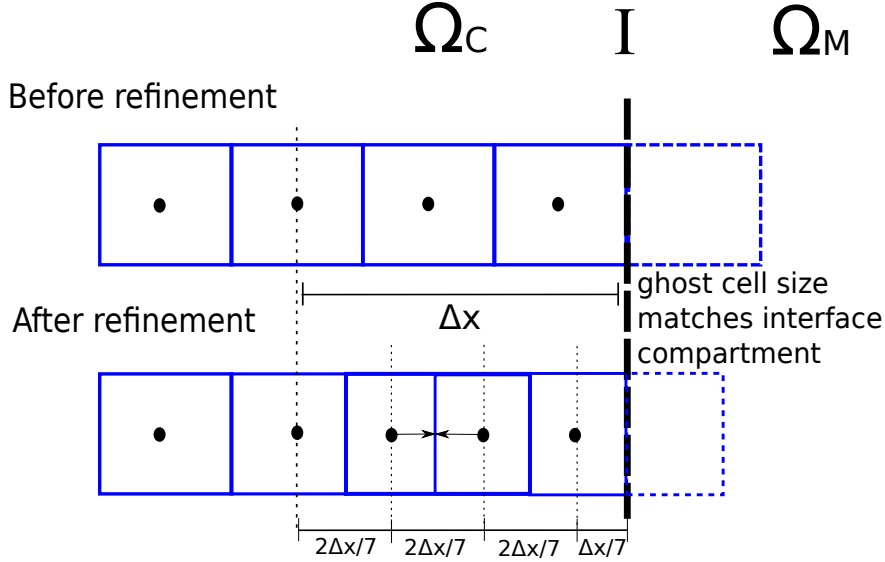}
\caption{{\it Diagram of one iteration of the lattice refinement} [R.1]--[R.3].}
\label{refinement}
\end{center}
\end{figure}

In order to see the effect of the compartment spacing near the interface 
$h$ on the error $E$ for both the TRM and GCM we start with the set of
regular compartments
$$
(0,h_0), \quad
(h_0, 2h_0), \quad
\dots,\quad
(0.5-h_0,0.5),
$$
which have nodes (centres of compartments) at $h_0/2$, $3h_0/2$, \dots,
$0.5-h_0/2$. Then we use the following lattice refinement technique designed 
specifically so that the position of the interface does not change
(see Figure \ref{refinement}):
\begin{enumerate}
 \item[\mbox{[R.1]}] Delete the two nodes closest to the interface.
 \item[\mbox{[R.2]}] Introduce into the space between the new node 
 closest to the interface and the interface (a distance of $\Delta x$) 
 three nodes placed consecutively a distance of $2\Delta x/7$ from the node to its left.
 \item[\mbox{[R.3]}] Recompute the compartments by finding the bisectors of 
 each node. 
\end{enumerate}
The specific distances in the step [R.2] are chosen such that the interface does 
not change location and the last two compartments have the same size. This 
is also the size that is given to the ghost cell in the GCM. The refinement algorithm 
[R.1]--[R.3] is repeated $m$ times such that the size 
of the final compartment in $\Omega_C$ (and ghost cell), $h_m$, is
\begin{equation}
 h_m = h_0\left( \frac{5}{7}\right)^m.
\end{equation}
A diagram representing one iteration of the refinement technique [R.1]--[R.3]
is shown in Figure \ref{refinement}.
The error is computed for various final compartment sizes $h_m$ ($m=0,1,\ldots,10$) 
and various time steps $\Delta t_k$ ($k=0,1,\ldots,10$) where
\begin{equation}
\Delta t_k =  2^k\Delta t_0,
\end{equation}
and $\Delta t_0 = 5\times 10^{-6}$.

\begin{figure}[t]
\begin{center}
 \includegraphics[scale=0.5]{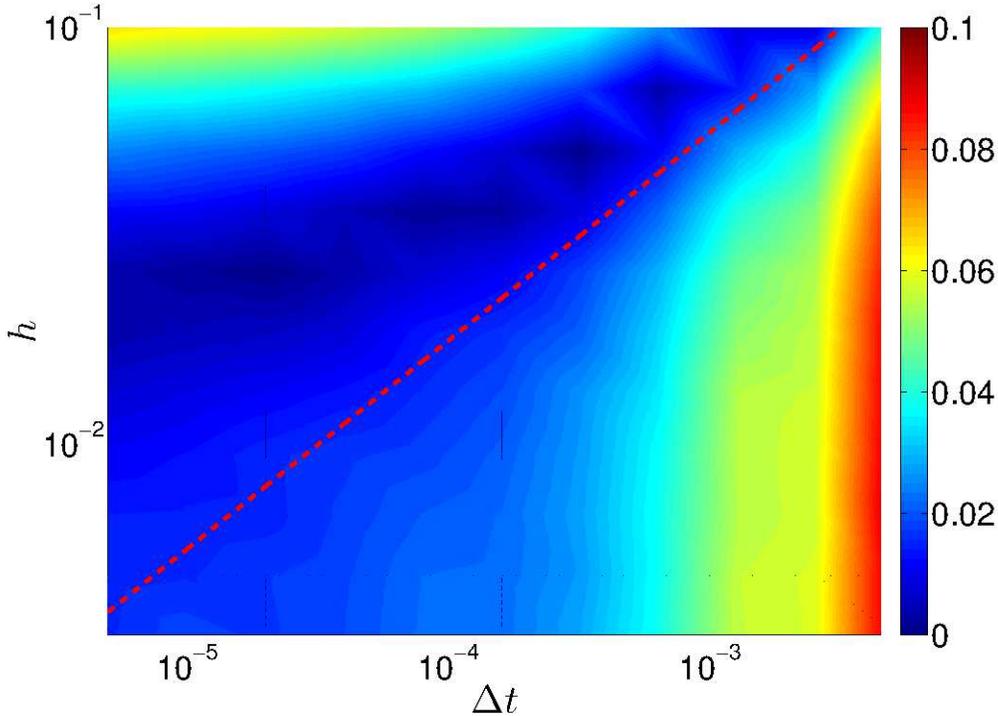}
\caption{{\it Surface plot of the absolute error $\left\|E\right\|$ defined by $(\ref{error})$ 
as a function of compartment size at the interface $h_m$ and time step $\Delta t$ for the one 
dimensional test problem using the TRM.}}
\label{TRMsurface}
\end{center}
\end{figure}

\begin{figure}[t]
\begin{center}
 \includegraphics[scale=0.5]{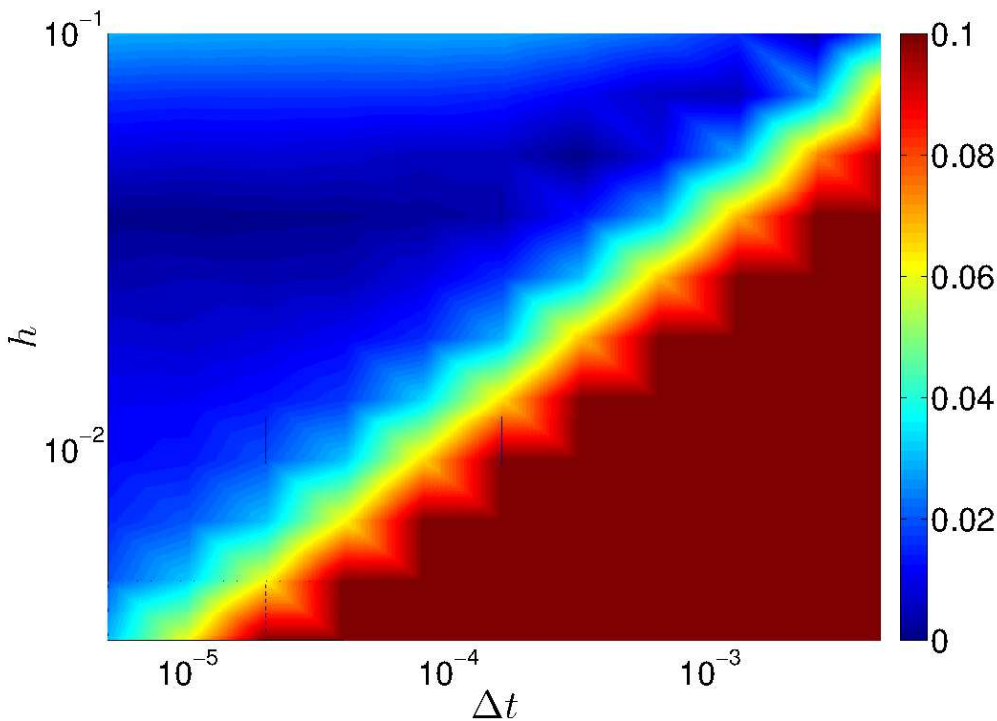}
\caption{{\it Surface plot of the absolute error $\left\|E\right\|$ defined by $(\ref{error})$ 
as a function of compartment size at the interface $h_m$ and time step $\Delta t$ for the one 
dimensional test problem using the GCM. }}
\label{GCMsurface}
\end{center}
\end{figure}

Figures \ref{TRMsurface} and \ref{GCMsurface} show how the absolute error $\left\|E\right\|$ 
given by (\ref{error}) depends on both parameters $h_m$ (compartment size on the interface) 
and $\Delta t$ for the TRM and GCM algorithms respectively. The error due to the interface 
in the TRM includes a shift of $h_m/2$ in the expected distribution 
of molecules at the interface into $\Omega_C$ \cite{Flegg:2012:TRM}. This is because of the 
``initialization'' 
of molecules from $\Omega_M$ into $\Omega_C$. Unlike the initialization of molecules 
from $\Omega_C$ into $\Omega_M$, molecules that are transported in the reverse direction 
cannot be placed carefully according to a continuous distribution but must necessarily be 
placed in the nearest compartment. This initialization has an expected position of $h_m/2$ 
away from the boundary causing a shift of $h_m/2$ in the distribution of molecules. However, 
if molecules could be initialized into $\Omega_C$ with a continuous distribution, for 
symmetry reasons one would expect this to be done with a distribution of $f(x)$ given by
(\ref{distribution}). The average distance, therefore, that a molecule would ideally
 be placed into $\Omega_C$ is $\int_0^\infty xf(x) dx = \sqrt{\pi D \Delta t}/2$. Therefore,
  the error that is due to unphysical shifting of molecules is proportional to the expected 
  shift of molecules as they are transferred from $\Omega_M$ to $\Omega_C$. That is 
  $E\propto h_m-\sqrt{\pi D \Delta t}$. 
  
\begin{figure}[t]
\begin{center}
 \includegraphics[scale=0.5]{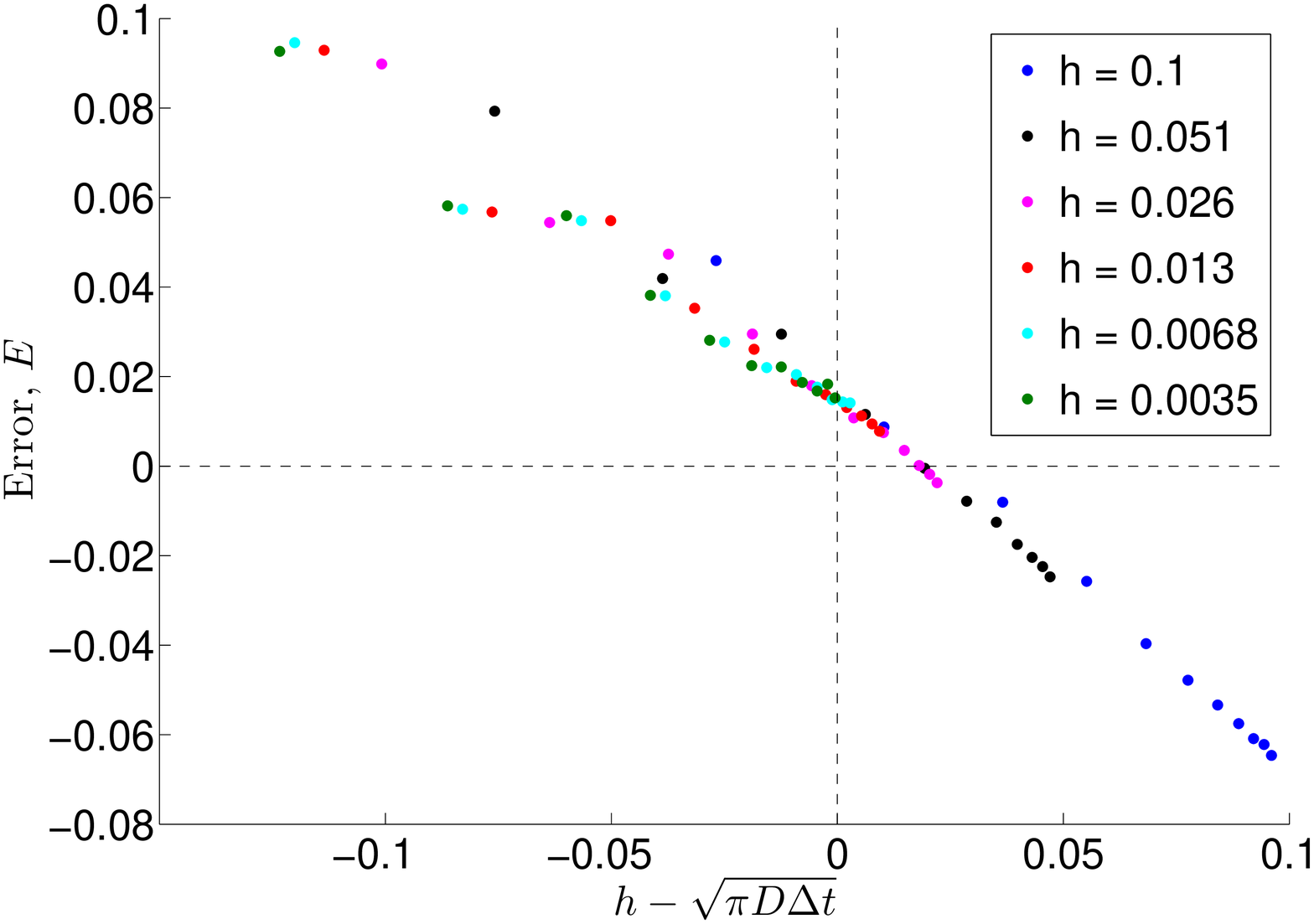}
\caption{{\it Scatter plot of the error $E$ versus $h_m-\sqrt{\pi D \Delta t}$ for the TRM. 
The different color points represent different values of the compartment size 
at the interface $h$ (see legend) and in each instance $\Delta t$ is varied from
 $5\times 10^{-6}$ to $5\times 10^{-3}$.}}
 \label{plotdataTRM}
\end{center}
\end{figure}

  In Figure \ref{TRMsurface} a dotted red line 
  showing $h_m=\sqrt{\pi D \Delta t}$ approximately follows the path of the minimum 
  absolute error. The discrepancy between the actual minimum absolute error and the 
  dotted red line in Figure \ref{TRMsurface} can be attributed to higher order error 
  that is inherent in the mesoscopic approximation to the diffusion equation. 
  To show that $E\propto h_m-\sqrt{\pi D \Delta t}$, Figure \ref{plotdataTRM} 
  is a plot of error $E$ versus  $h_m-\sqrt{\pi D \Delta t}$. The plot is generated by using
   various values of $h_m$ (see legend) and then plotting a number of points while changing 
   $\Delta t$. Whilst it is clear that the graph is approximately linear, the higher order 
   mesoscopic error is clearly seen in the form of a vertical displacement of this curve 
   about the origin. The effect that the higher order mesoscopic error has on the interface
    is difficult to quantify because it will depend on the particular molecular system. 
    Therefore, the best choice of parameters that can be chosen for the TRM is 
    $h_m\sim\sqrt{\pi D \Delta t}$. 
    
    In Figure \ref{GCMsurface}, we see that the error of the GCM depends 
    on $\Delta t$ and specifically on its relative size compared to 
    $h$ (the analysis of the GCM is provided in \ref{GCMproof}). Rapidly increasing 
    error (quickly saturating the color bar in Figure \ref{GCMsurface}) is observed 
    when $h_m \sim \sqrt{\pi D \Delta t}$. The higher order mesoscopic error artefact 
    can also be seen in Figure \ref{GCMsurface} since this artefact is independent 
    of the coupling mechanism (see the larger absolute error for large values of $h$). 
    The GCM is therefore most accurate for very small values of $\Delta t$. Whilst 
    in practice making $\Delta t$ small may significantly increase the computing time, 
    small $\Delta t$ is often required for accurate microscopic simulation (for example, 
    capturing reactions with high resolution) and in such cases the GCM is more appropriate 
    than the TRM.

\subsection{Three dimensional simulations: CPM versus GCM}

\noindent
In this section we will demonstrate how, when using an unstructured mesh, the error 
associated with the GCM coupling converges as $\Delta t\rightarrow 0$ whereas error 
associated with the CPM is minimized when $h \sim \sqrt{D\Delta t}$ where
$h$ is the average size of boundary compartments. Both the error 
associated with the CPM and GCM are due to imbalances in the flux of molecules over 
the interface. We implement the CPM and GCM in three spatial dimensions
using a tetrahedral primal mesh as described in Section \ref{URDME}. The implementation builds on the freely available software URDME \cite{Drawert:2012:URDME}.

\begin{figure}
\centerline{
\hskip 3mm
\raise 6.11 cm \hbox{(a)}
\hskip -8mm
\epsfig{file=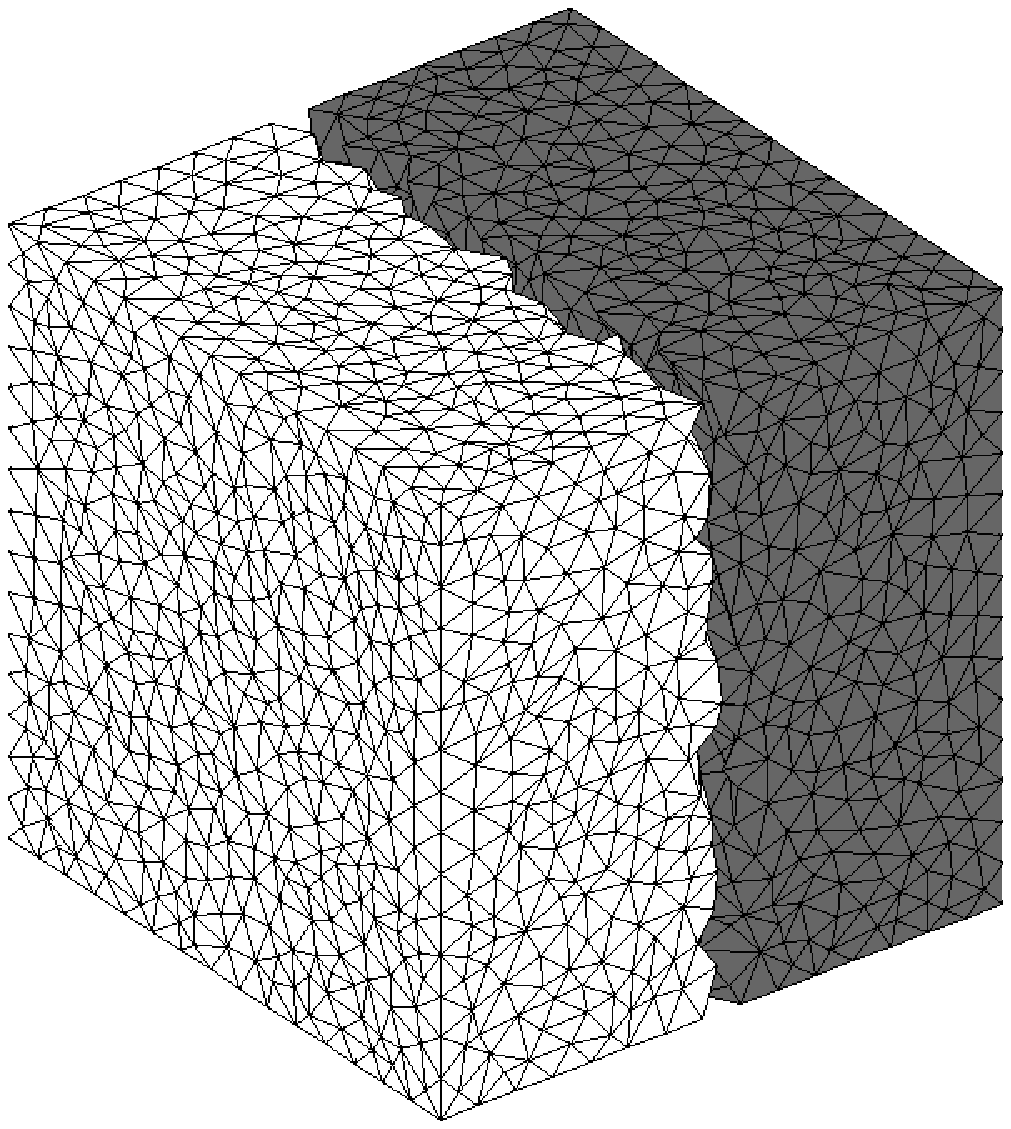,height=6cm}
\hskip 6mm
\raise 6.11 cm \hbox{(b)}
\hskip -8mm
\epsfig{file=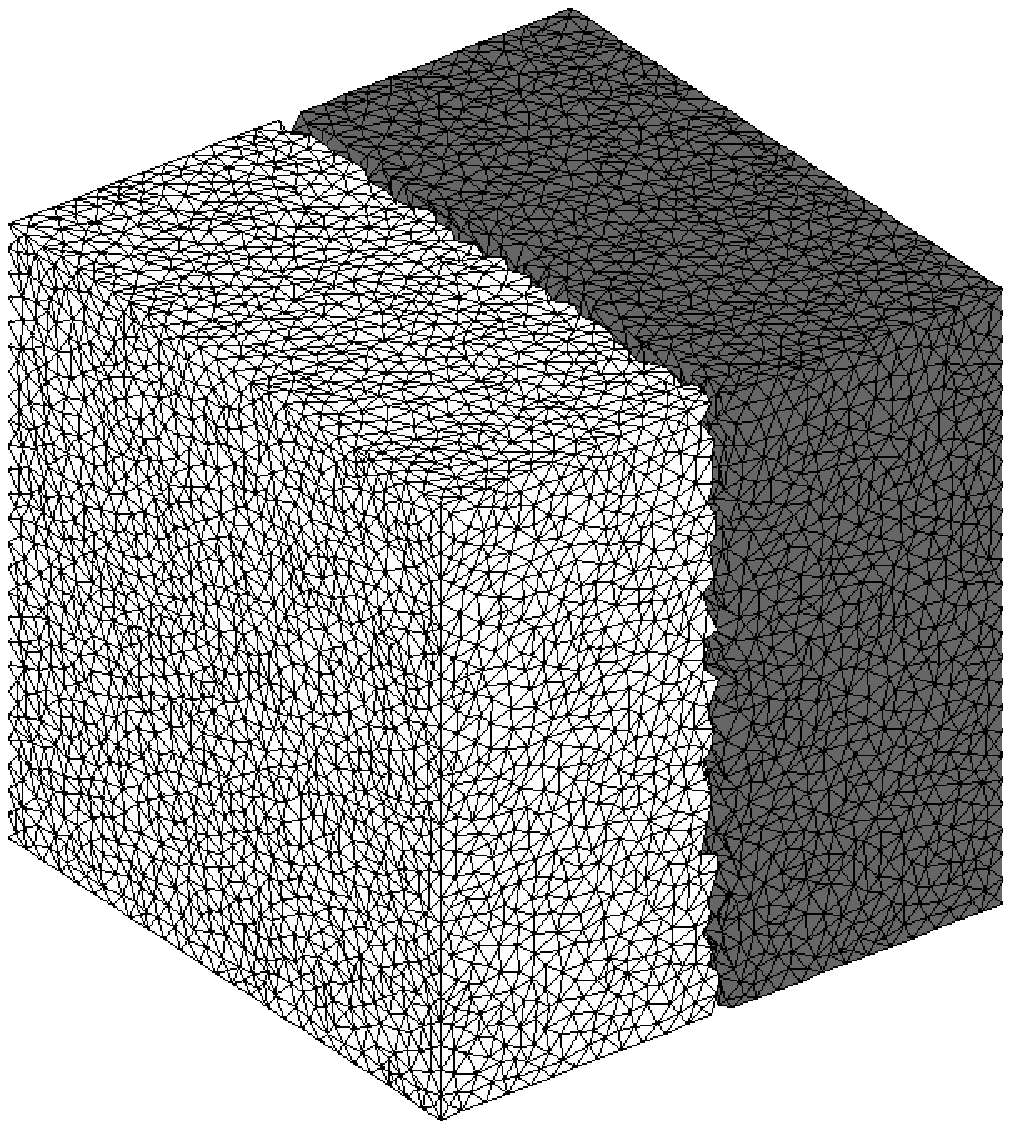,height=6cm}
}
\caption{{\it The cube $[0,1]^3$ is partitioned into a mesoscopic region (grey) and 
a microscopic region (white).} (a) {\it a coarse mesh}; (b) {\it a fine mesh}.}
\label{fig:partitioned_cube}
\end{figure}

\begin{figure}
\centering
\includegraphics[scale=0.65]{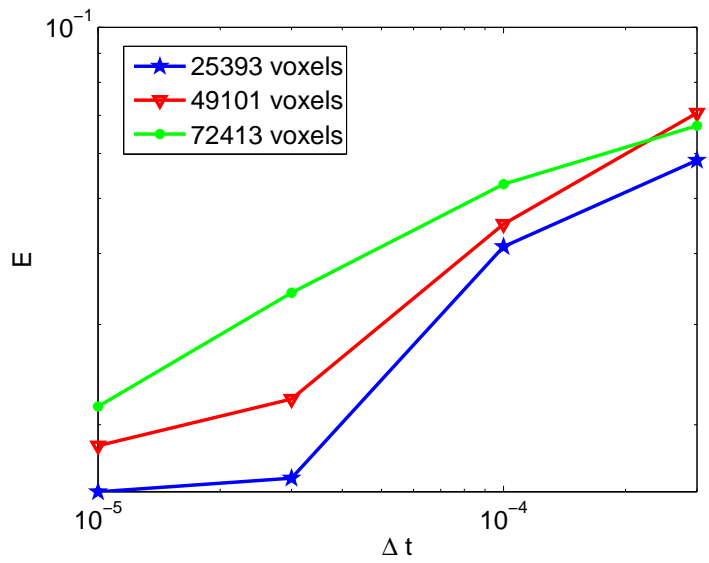}
\caption{{\it The error $\left\|E\right\|$ of the GCM method for different mesh 
sizes and time steps. The error decreases with decreasing time step, as expected.}}
\label{fig:results_3D}
\end{figure}

 We consider a cube with side length $L = 1$. The cube is first discretized 
 with an unstructured mesh and then divided into a mesoscopic region $\Omega_{\mathrm{C}}$, 
 and a microscopic region $\Omega_{\mathrm{M}}$, where $\Omega_{\mathrm{M}}$ is the 
 set of all voxels with a vertex $(x,y,z)$ such that $x<0.5$ and 
 $\Omega_{\mathrm{C}}=\Omega\setminus\Omega_{\mathrm{M}}$. Here $\Omega$ is the set 
 of all voxels. The partitioning is illustrated in Figure \ref{fig:partitioned_cube} 
 for two different mesh sizes.%
We start each simulation with $N_0 = 2\cdot 10^4$ molecules whose initial positions 
are sampled from a uniform distribution. The diffusion constant of the molecules is 
$D=1$, and we simulate the system for time t=$0.1$. Since we start with 
a uniform distribution and the molecules only diffuse and do not react, 
we expect the distribution to be uniform at the final time. As the interface 
is parallel with the $y-z$-plane, we expect that the distributions of molecules 
in the $y$- and $z$-directions are uniform, but that we get a small error 
in the distribution of molecules in the $x$-direction. We now divide the 
$x$-axis into $10$ bins of equal length, and then count the number of molecules 
in each bin at the final time. Mesoscopic molecules are binned by first sampling 
a continuous position from a uniform distribution on the voxel. We expect $N_0/10$ 
molecules in each bin, and can therefore estimate the error $E$ by
\begin{align}
\label{eq:error_cube}
E = \frac{\sum_{i=1}^{10} |N_i-N_0/10|}{N_0}.
\end{align}
In Figure \ref{fig:results_3D} we have computed $\left\|E\right\|$ for different 
mesh sizes and time steps. As expected, the error decreases as we refine the mesh 
and decrease the time step.

In the CPM method, mesoscopic (resp. microscopic) molecules stay mesoscopic 
(resp. microscopic) during a time step. This implies that the time step should 
be chosen sufficiently small such that a molecule does not diffuse across 
several voxels. On the other hand, if the time step is too small the distribution 
of molecules in space will be biased towards the microscopic region. This can 
be seen by considering a microscopic molecule diffusing into the mesoscopic regime. 
If the time step is small, it is likely that it will be close to the microscopic 
regime at the end of the time step, but if it ends up on the mesoscopic side 
it will nevertheless be considered uniformly distributed in the voxel at the 
end of the time step. Thus, the time step should not be chosen too small relative 
to the size of the voxels, or the error due to the spatial splitting will become 
large.

\begin{figure}
\centering
\includegraphics[scale=0.65]{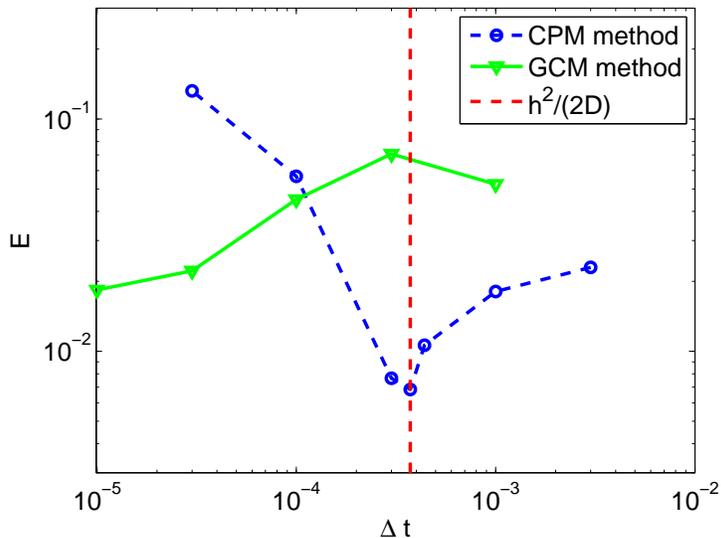}
\caption{{\it Comparison of the error $\left\|E\right\|$ produced by the GCM and CPM 
for interfacing microscopic and mesoscopic simulations as a function of the time step 
in the microscopic simulation domain $\Delta t$. The length scale $h$ is defined to be 
the cubic root of the average volume of a voxel.}}
\label{fig:comparison}
\end{figure}

Since the GCM converges with decreasing time step, but performs worse for larger 
time steps, one could suspect that there is a regime where the CPM 
in \cite{Hellander:2012:CMM} performs better than the GCM. At some point, 
however, the error of the GCM will become small and outperform the CPM
in \cite{Hellander:2012:CMM}. The errors of the different methods are compared 
in Figure \ref{fig:comparison} for a mesh with $49101$ voxels. Indeed, 
we see that the CPM method performs better for time steps down to almost 
$\Delta t=10^{-4}$, at which point the error of the GCM method becomes smaller.

\section{Summary}
\label{conc}

\noindent
In this paper we have compared two existing mesoscopic-microscopic coupling 
techniques for stochastic simulations of reaction-diffusion processes with 
a new convergent method called the ghost cell method (GCM). Here we will 
summarize the specific sources of error of the TRM, CPM and GCM, when they 
converge, how they may be optimized for accuracy and notes on their computational 
efficiency.

\subsection{Summary of the two-regime method}

\noindent
The TRM couples molecules by considering that mesoscopic compartments 
contain molecules that are evenly distributed in a probabilistic sense. 
As these molecules diffuse over the interface they are placed according 
to the distribution $f(x)$ given by (\ref{distribution}). Molecules 
migrating in reverse from the microscopic regime to the mesoscopic 
regime must be absorbed by the interfacial compartments and be indistinguishable 
from other molecules in these compartments by virtue of this paradigm. 
As such, instead of migrating an average distance over the interface 
proportional to $\sqrt{\Delta t}$ it becomes evenly distributed over 
the compartment with an expected location of $h/2$ over the interface. 
The molecules are therefore effectively shifted $(h-\sqrt{\pi D\Delta t})/2$ 
into the compartment regime. This shift in the molecules therefore creates 
a discontinuity of in the distribution to find molecules on the interface 
and therefore an error due to the presence of the coupling proportional 
to $h-\sqrt{\pi D \Delta t}$. The nature of this error is that, if the 
expected net flux of molecules over the interface is 0, then no error 
due to the presence of the interface will be experienced. This is important 
to note, since, this is not the case for both the CPM and GCM methods.
Furthermore, since the error is proportional to $h-\sqrt{\pi D \Delta t}$ 
it clearly converges in the limiting case (ii) described in the introduction 
but not in the limiting case (i). 

Whilst the TRM can give controllably accurate results, it can be computationally 
more costly to implement. This is because perfect absorption of molecules 
is required on the boundary and this means that each molecule in the 
molecular-based domain needs to be checked for interaction with the boundary 
in a given time step using (\ref{PM}).

\subsection{Summary of the compartment-placement method}

\noindent
The CPM is a coupling mechanism that, whilst heuristically derived, can 
produce accurate results under some circumstances and do so with minimum 
computational cost. Molecules are placed within a pseudo-compartment 
in the molecular-based domain via diffusion from the compartment-based domain. In reverse 
molecules are placed in compartments from the molecular-based domain via diffusion 
of these molecules in the continuous domain over the interface. The antisymmetry 
that is seen in the methods of migration, mesoscopic to microscopic and microscopic 
to mesoscopic, result in a boundary layer in the expected distribution of molecules 
at the interface. This boundary layer is caused by the fact that molecules diffusing from the molecular region to the compartmental region are considered uniformly distributed on the compartment at the end of the time step. If $\Delta t$ is small compared to $h$, this will be a poor approximation. It should thus be noted that the error of the 
CPM does not converge in the limiting case (i) described in the introduction, however 
the error appears to converge according to limiting case (ii).

The CPM is computationally efficient. Its only inefficiency is that, unlike the TRM, 
it requires the knowledge of a pseudo-compartment in the microscopic domain. The TRM 
is therefore more appropriate than the CPM for coupling completely independent 
simulation algorithms, since the CPM requires its own custom algorithm to 
be implemented fully.

\subsection{Summary of the ghost cell method}

\noindent
The GCM couples molecules according to a discrete 
domain on each side of the interface. Molecules that are in the microscopic domain are
 binned according to a ghost compartment/cell and jump into the mesoscopic domain using 
 jump rates derived using the mesoscopic approach. In such a way, symmetry is conserved in the method 
 of migration from mesoscopic to microscopic and from microscopic to mesoscopic, unlike the CPM. 
 It is important that the molecules are binned correctly for this coupling to work accurately. 
 The error, therefore, can be attributed to molecules that are in the ghost cell when they 
 should not be, or not in the ghost cell when they should be. Therefore, if the compartment 
 size $h$ at the interface (and of the ghost cell) is much larger than the resolution of the 
 particle tracking in the microscopic domain, the correct number of molecules will be in the ghost 
 cell. The error therefore converges in the limit of small $\Delta t$ so 
 long as $h$ is sufficiently coarse. Furthermore, it is possible to show that, unlike the TRM, 
 this source of error is not due to a displacement of molecules but an unballanced flux 
 of molecules (see \ref{GCMproof}) and will therefore appear even if the expected net flux 
 over the interface is 0. The GCM, however, is convergent in the limiting case (i) but not 
 (ii) from the introduction giving the GCM a unique advantage over both the CPM and TRM.

The GCM is computationally efficient for small $\Delta t$ since the jump rates from the 
microscopic domain to the mesoscopic domain are determined by the ghost cell size and not 
the time step (like the TRM for example). 


\vskip 1cm

\noindent {\bf Acknowledgements:} \hskip -0.07mm The research leading to these
results  has received funding from the \hbox{European} Research Council under
the  {\it European Community}'s Seventh Framework Programme ({\it
FP7/2007-2013})/ ERC {\it grant agreement} No. 239870.  
This publication was based on work supported in part by Award 
No KUK-C1-013-04, made by King Abdullah University of Science 
and Technology (KAUST). 
Stefan Hellander has been supported by the National Institute of Health 
under Award no. 1R01EB014877-01 and the Swedish Research Council. 
Radek Erban would also like to thank Brasenose College, University of Oxford, 
for a Nicholas Kurti Junior Fellowship; the Royal Society for a University 
Research Fellowship; and the Leverhulme Trust for a Philip Leverhulme Prize. 

\appendix

\section{Mathematical justification for the ghost cell method}
\label{GCMproof}

\noindent
Here we present an analysis of the GCM in one-dimension. We show 
that error of the GCM that is produced on the interface between mesoscopic 
and microscopic subdomains converges in the case (i). Specifically, 
we see convergence of the interface-derived error as 
$\Lambda = \sqrt{D\tstep}/h \rightarrow 0$. This property of 
convergence is unique to the GCM when compared with other 
reported coupling mechanisms. In showing that the interface-derived 
error vanishes in the small time step limit, we will show that 
rapid variation within the boundary layer of the interface 
vanishes as $\Lambda \rightarrow 0$, leaving behind a linear 
approximation of the true distribution of molecules. Since the 
error at the interface will be of order $h^2$ it is accurate to 
the same order as the mesoscopic algorithm itself.
 
Without loss of generality, consider an interface at $x=0$ on an 
infinite one-dimensional domain. To the left of this interface ($x<0$) 
a compartment-based model is used with fixed compartment size $h$. 
To the right of the interface ($x>0$) a molecular-based algorithm 
is used and is updated at fixed time increments of $\tstep$,
i.e. $\Omega_C = (-\infty,0)$ and $\Omega_M = (0,\infty)$.
We denote the compartments in $\Omega_C$ by $C_k = (-kh,-kh+h)$, where
$k = 1,2,\dots.$ Then the interface compartment is $C_1 = (-h,0)$. 
The ghost cell will be denoted by $C_M = (0,h)$.

Molecules in $\Omega_C$ are described only by their compartment. 
Their compartment changes with an exponentially distributed random 
time with a rate that is given by $D/h^2$. This rate is conditional 
on initial and final states being compartments. The rate given 
by $D/h^2$ is chosen in such a way that the expected number of 
molecules in each compartment matches that of a discretized diffusion 
equation (see (\ref{equationformean}) and (\ref{equationforq})). 
These rates, however, breakdown in the case of the TRM because 
the initial and final states of jump across the interface are not 
compartments but rather the final state is a molecule in $\Omega_M$.
The jump across the boundary for the TRM is given by (\ref{Gamma}). 
Molecules in $\Omega_M$ have one thing in common with those in compartments. 
A domain that is modeled microscopically and then binned into compartments 
shows the same expected behaviour as a domain modeled with compartments 
to leading and first order accuracy in $h$ in the limit as 
$\Delta t \rightarrow 0$. Therefore, in an attempt to interface 
the two regimes together it may be appropriate to bin  
molecules in $\Omega_M$ into a ghost cell/compartment $C_M$
near the interface. The molecules that are in $C_M$ will then 
have the same properties as a compartment from the perspective 
of the interface compartment $C_1$. To this end, any molecule in 
$C_M$ may spontaneously change state from the molecular domain 
to $C_1$. We expect that since the interface compartment-bound 
molecules see a compartment state for molecules in $C_M$, 
the change of state from $C_1$ to a random position within 
$C_M$ will occur with a normal inter-compartmental rate. 
In the following analysis we show that this is the case.

We shall test the hypothesis by matching the master equations 
for $C_1$ and the probability distribution 
in $\Omega_M$ in such a way that no rapid variation in probability 
to find molecules, $\bar{p}(x,t)$, is apparent at the interface. 
We shall assume that the rate for molecules to jump into 
$\Omega_M$ from $C_1$ is $\Gamma^+$ and are placed 
in an initial position from the interface given by the 
probability distribution $f(x)$. Molecules in $\Omega_M$ 
spontaneously jump into $C_1$ with 
a rate $\Gamma^- g(x)$. Functions $f(x)$ and $g(x)$ are 
normalized such that they have a unit integral over
$\Omega_M$. We shall show that 
\begin{equation}
\Gamma^+ = \Gamma^- = \frac{D}{h^2}
\qquad
\mbox{and} 
\qquad
g(x) = f(x) = \left\{ \begin{matrix}
\displaystyle\frac{1}{h}, & \mbox{for} \; x \in C_M; \\
\rule{0pt}{5mm} 0, & \mbox{otherwise.} 
\end{matrix}\right.
\label{GCMresult}
\end{equation}
We will find it convenient for the sake of notation to 
introduce the parameters
$$
\alpha^\pm = \frac{h^2\Gamma^\pm}{D},
\qquad
\mbox{and}
\qquad
\Lambda = \frac{\sqrt{D\tstep}}{h}.
$$ 
To show (\ref{GCMresult}) we focus on the purely diffusive
problem, since bulk reactions have no effect on boundary conditions. 
In order to limit the flux of molecules jumping into $C_1$, 
all molecules in $\Omega_M$ that hit the interface 
by Brownian motion are reflected back to $\Omega_M$.

We denote  the probability of finding a molecule in compartment
$C_k$, $k=1,2,\dots,$ by  $p_k(t)h$ (so that $p_k(t)$ approximates the
probability density function at the node within this compartment). 
If we denote by $p(x,t)$ the probability density function of the 
discrete-time molecular-based algorithm, then the 
transmission/reflection rules give us the following master equation
\begin{eqnarray}
\label{master1}
p_1(t+\tstep) &=& \left( 1 - (1+\alpha^+) \Lambda^2 \right) p_1(t) +
\Lambda^2 \, p_{2}(t) 
+  \alpha^- \Lambda^2 \int_0^\infty g(x)p(x,t) \,\mbox{d}x, 
\qquad \\
\label{master2}
p(x,t+\tstep) &=& \int_0^\infty \frac{p(y,t)}{\sqrt{4\pi D
    \tstep}}\left[ \exp\left( \frac{-(x-y)^2}{4D\tstep}\right)
 +\exp\left( \frac{-(x+y)^2}{4D\tstep}\right)\right]\, \mbox{d}y  
 \nonumber  \\ && \mbox{ }+ \frac{\alpha^+D\tstep p_{1}(t)f(x)}{h} 
 - \frac{\alpha^-D\tstep p(x,t)g(x)}{h} .
\end{eqnarray}
In the vicinity of $x=0$ there is a boundary layer
of width $O(h)$ so long as $f(x)$ and $g(x)$ vanish for $x \gg h$
\cite{Erban:2007:RBC}. 
We rescale (\ref{master1}) and (\ref{master2}) using the
(dimensionless) boundary 
layer coordinate $\xi = x/h$. 
We also denote $p_{\mathrm{inner}}(\xi,t) =
p(h\xi,t)$, $f_{\mathrm{inner}}(\xi)= h\,
f(h\xi)$ and $g_{\mathrm{inner}}(\xi)= h\,
g(h\xi)$. The rescalings of $f$ and $g$ by $h$ are done to keep 
the integrals of these functions equal to 1.
Thus, in the boundary layer coordinates, (\ref{master1}) and
(\ref{master2}) become 
\begin{eqnarray}
\label{master3}
p_1(t+\tstep) &=& \left( 1 - (1+\alpha^+)\Lambda^2 \right) p_1(t) +
\Lambda^2 p_{2}(t) \nonumber \\ && \mbox{ }
+ \alpha^- \Lambda^2 \int_0^\infty g_{\mathrm{inner}}(\xi)
\, p_{\mathrm{inner}}(\xi,t)\, \mbox{d}\xi, \\
\label{master4}
p_{\mathrm{inner}}(\xi,t+\tstep) &=& 
\Lambda^{-1}\int_0^\infty p_{\mathrm{inner}}(\eta,t)
\left[ K\left(\Lambda^{-1}(\eta-\xi)\right) 
+ K\left(\Lambda^{-1}(\eta+\xi)\right)\right]\, \mbox{d}\eta  \nonumber  
\\ && \mbox{ }+ \alpha^+\Lambda^2 p_{1}(t)\, f_{\mathrm{inner}}(\xi) 
- \alpha^-\Lambda^2 p_{\mathrm{inner}}(\xi,t)\, g_{\mathrm{inner}}(\xi),
\end{eqnarray}
where $K(x) = (4\pi)^{-1/2} \exp(-x^2/4)$.
The parameter $\Lambda$ gives us the relative size of $D \Delta t$ to 
$h^2$ and we wish to show that as $\Lambda\rightarrow 0$ the distribution 
of molecules across the boundary is smooth and the error that remains 
is of the order of $h^2$, which is the same size of the error associated 
with the mesoscopic discretization in $\Omega_C$. In order to join 
these models smoothly we require in $\Omega_C$ that 
\begin{eqnarray}
p_1(t) &=& p(-h/2,t) = p(0,t) - \frac{h}{2}\bar{p}_{x}(0,t) + O(h^2) 
+ O(\Lambda), \label{condition2} \\ 
p_{2}(t) &=& p(-3h/2,t) = p(0,t) - \frac{3h}{2}\bar{p}_{x}(0,t) 
+ O(h^2) + O(\Lambda), \\
p_1(t+\Delta t) &=& p_1(t) + O(\Delta t), 
\end{eqnarray}
while, for the molecular-based side, we want variation from the 
linear approximation in the
boundary layer to be limited to $O(\Lambda)$ up to order $h^2$ accuracy, 
so that
\begin{eqnarray}
p_{\mathrm{inner}}(\xi,t) &=& p(0,t) + h\,\xi\,
p_x(0,t) + O(h^2) + O(\Lambda), \\
p_{\mathrm{inner}}(\xi,t+\Delta t) &=& 
p_{\mathrm{inner}}(\xi,t) + O(\Delta t). \label{expansion}
\end{eqnarray}
The prescription of a consistent probability density $p(0,t)$ 
and derivative $p_x(0,t)$ for both sides of the interface, 
along with linear approximations sufficiently close to the 
interface equates to continuity and differentiability over the 
interface which is the matching condition that we are attempting 
to achieve. 

Substituting (\ref{condition2})--(\ref{expansion}) into 
(\ref{master3}) and (\ref{master4}) and equating terms 
of the same order in $h$ and leading order in $\Lambda$ 
gives the following conditions that must be placed on 
$g(x)$, $f(x)$, $\alpha^+$ and $\alpha^-$:

\vskip 2mm

{\leftskip 8mm

\parindent -5mm

(i) $O(h^0\Lambda^0)$ terms from equation (\ref{master3}) give condition:
\begin{equation}
\label{condition1a}
\alpha^+  =
 \alpha^-\int_0^\infty g_{\mathrm{inner}}(\xi), \mbox{d}\xi .
\end{equation}
Condition (\ref{condition1a}) 
states how the relative rates for molecules to transition to 
and from $\Omega_M$ and $\Omega_C$ must be dependent on the 
relative sizes of $C_1$ and $C_M$. 

\vskip 2mm

\parindent -6.5mm

(ii) $O(h^1\Lambda^0)$ terms from equation (\ref{master3}) give condition:
\begin{equation}
\label{condition2a}
  2 = \alpha^+ + \alpha^- \int_0^\infty \xi g_{\mathrm{inner}}(\xi), \mbox{d}\xi .
\end{equation}
Condition (\ref{condition2a}) 
states how the rates for molecules to transition to and from 
$\Omega_M$ and $\Omega_C$ depend on the average distance molecules
 are placed from the interface when placed within $C_M$. 
 This is the same condition given in $\Omega_C$ 
 for jumps between the compartments.   

\vskip 2mm

\parindent -8mm

(iii) $O(h^0\Lambda^0)$ terms from equation (\ref{master4}) give condition:
\begin{equation}
\label{condition3a}
  \alpha^+ f_{\mathrm{inner}}(\xi) = \alpha^- g_{\mathrm{inner}}(\xi).
\end{equation}
Condition (\ref{condition3a}) states that molecules must 
be placed into $\Omega_M$ with the same probability weighting 
that they are taken out and placed back into $\Omega_C$. 

\vskip 2mm

\parindent -7mm

(iv) $O(h^1\Lambda^0)$ terms from equation (\ref{master4}) are automatically 
satisfied. 

\leftskip 0cm
\parindent 0cm
}

\vskip 2mm

\noindent
The GCM that is presented in this manuscript uses parameters
(\ref{GCMresult}) which satisfy 
the three conditions (\ref{condition1a})--(\ref{condition3a})
listed above. 
Such a scheme, therefore, has an error that is no greater 
than the error of the mesoscopic scheme in the limit 
$\Lambda \rightarrow 0$, in other words, in the 
limit $\Delta t \rightarrow 0$ whilst $h$ remains constant.


\end{document}